# Coexistence of fast photodarkening and slow photobleaching in $Ge_{19}As_{21}Se_{60}$ thin films

Pritam Khan, [1] A. R. Barik, [1] E. M. Vinod, [2] K. S. Sangunni, [2] H. Jain, [3] and K. V. Adarsh [1]

[1] Department of Physics, Indian Institute of Science Education and Research, Bhopal 462023, India
[2] Department of Physics, Indian Institute of Science, Bangalore 560 012, India
[3] Department of Materials Science and Engineering, Lehigh University, Bethlehem, Pennsylvania 18015, USA
pritam@iiserb.ac.in

**Abstract:** We experimentally demonstrate the coexistence of two opposite photo-effects, viz. fast photodarkening (PD) and slow photobleaching (PB) in $Ge_{19}As_{21}Se_{60}$ thin films, when illuminated with a laser of wavelength 671nm, PD appears to begin instantaneously upon light illumination and saturates in tens of seconds. By comparison, PB is a slower process that starts only after PD has saturated. Although we could observe the coexistence of PD/PB even at moderate, one order of magnitude lower intensity of 0.2 W/cm$^2$, the kinetics of transformation is significantly slowed down. However, both PD and PB follow stretched exponetial dependence on time. Modeling of overall change as a linear sum of two contributions suggests that the changes in As and Ge parts of glass network respond to light indepndent of each other.

## 1. Introduction

Chalcogenide glasses (CG) exhibit numerous photoinduced effects with bandgap or sub-bandgap illumination. Among them the most notable effects are photodarkening (PD) in As-based chalcogenides and photobleaching (PB) in Ge-based chalcogenides [1, 2]. Besides the fundamental interest in PD/PB in CG, these effects find useful technological applications in high bit rate waveguide writing, dense holographic recording, etc. [3-6]. PD in As-based CG has been investigated for many years and is believed to originate from photoinduced structural transformations. Interestingly, PD consists of a transient and a metastable part [7, 8], such that the former decays quickly once the illumination is switched off, leaving behind the metastable part that can only be reversed by annealing near the glass transition temperature [2]. By contrast, light illumination in Ge-based CG mostly results in PB and it appears to be irreversible [2]. However, recent studies on $GeSe_2$ thin films have shown that the irreversible PB is accompanied by a transient PD [9, 10]. Many models have been proposed to explain the observed PB in Ge-based CG, but none of them appears to be applicable for all compositions [11, 12]. In general, it is believed that intrinsic structural changes and photo-oxidation are responsible for the observed PB in Ge-based CG [10]. Naturally, PD in As based and PB in Ge based CG glasses calls for experiments to observe the light-induced behaviour of a glassy system consisting of Ge, As and Se as major constituents and to determine how the associated As-Se and Ge-Se parts respond to light illumination jointly. Such information will provide new insight in understanding PD/PB in CG. With this motivation, we have investigated photo-induced changes in the optical transmittance of $Ge_{19}As_{21}Se_{60}$ thin films. We demonstrate that light illumination can induce an unusual coexistence of fast photodarkening (PD) and slow photobleaching (PB). In addition, we found that PD appears to begin instantaneously with light illumination and saturates immediately, however on the other hand PB is a slow process and takes a longer time to saturate. Although we could observe the coexistence of PD/PB even at moderate, one order of magnitude lower intensity of 0.2 W/cm$^2$, the kinetics of transformation is significantly slowed down.

## 2. Experimental details

Bulk $Ge_{19}As_{21}Se_{60}$ glass was prepared by melt-quench method using 99.999% pure As, Ge and Se powders. The cast sample was used as the source material for deposition, and thin film of average thickness ~1.0 μm was prepared by thermal evaporation in a vacuum of about $1 \times 10^{-6}$ Torr. PD/PB in these films was studied by a pump-probe optical absorption method using the experimental set up described previously [13]. We have chosen the wavelength of the pump beam as $\lambda = 671$nm (from a diode pumped solid state laser, DPSSL) and kept its intensity at 2W/cm$^2$. The probe beam was a low intensity white light in the wavelength regime of 450-1000 nm. During illumination, transmission of the sample was recorded using Ocean Optics high resolution composite grating spectrometer (HR 4000 CG), which has the capability to collect the entire optical pectrum in 2ms. In our experiments, full optical spectrum was collected in real time of 100 ms/spectrum.

## 3. Results and discussion

To estimate the effect of pump beam illumination, first we recorded the transmission spectrum of as-prepared sample in dark condition and denoted it as $T_i$. Next, we turned on the pump beam and continuously recorded the transmission spectrum as a function of time ($T_f$) until the whole effect saturated, which took nearly an hour. Fig. 1 shows the contour plot of $T_f/T_i$ as a function of time on selected wavelengths close to the optical bandgap of the sample. It is clear from the fig. that PD appears to begin instantaneously after turning on the pump beam and saturates within a few hundreds of seconds. To our great surprise, after complete saturation of PD, PB starts to develop and grows, showing a remarkable shift in the transmission. Finally, it also saturates at a value which is well above the initial value. To understand the role of intensity on the unusual coexistence of PD and PB, we have performed the same experiment at a moderately low intensity of 0.2 W/cm$^2$. Basically, the whole effect is reproducible even at such low intensities. These experimental results clearly demonstrate the unusual coexistence of PD and PB in $Ge_{19}As_{21}Se_{60}$ thin films.

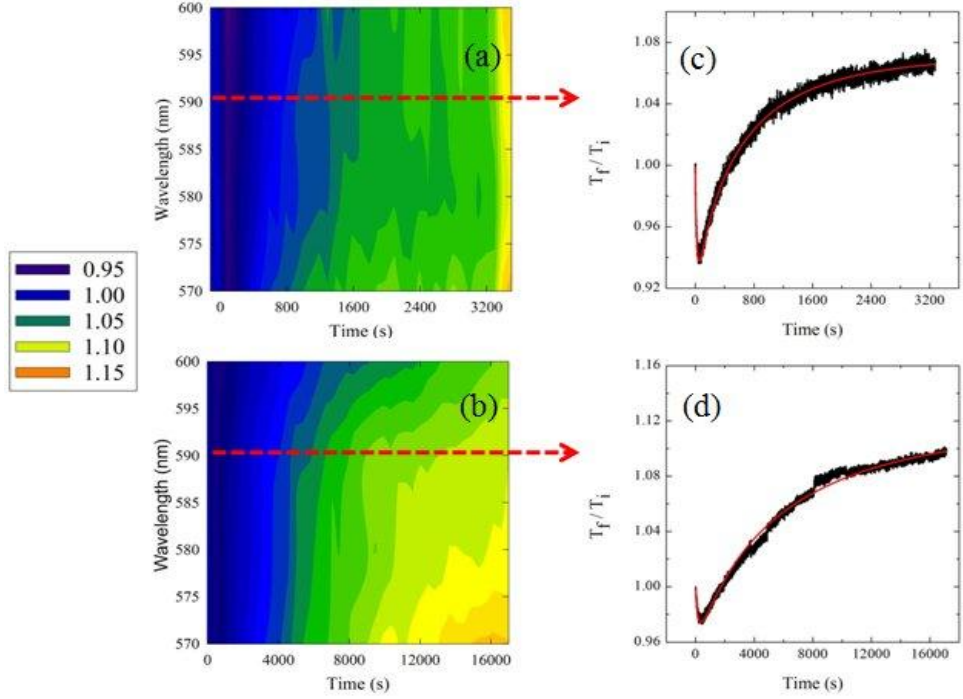

**Fig. 1.** Contour plot of $T_f/T_i$ as a function of time at selected wavelengths close to the optical bandgap of $Ge_{19}As_{21}Se_{60}$ thin film when irradiated with a 671nm laser of intensity (a) 2W/cm$^2$ and (b) 0.2W/cm$^2$. The color bars indicate the value of $T_f/T_i$, where $T_f$ and $T_i$ are transmission spectrum of the sample at a particular time after the laser was turned on and for as-prepared sample in dark condition respectively. At indicated wavelengths, transmission decreases initially, followed by a sharp increase in transmission. Time evolution of $T_f/T_i$ at the wavelength 590nm for the pump beam intensity (c) 2W/cm$^2$ and (d) 0.2W/cm$^2$, here the black and red line represent the experimental data and theoretical fit (using eq. 5), respectively. Figures clearly demonstrate the coexistence of PD and PB with the former being a significantly faster process.

To explain the unusual coexistence of PB and PD on the same sample, we assume that the there exist compositional heterogeneities (As-As, Ge-Ge etc ) in the film, which were created from the vapour that contains various atomic fragments [14,15]. When such a film is illuminated with 671nm light, compositional heterogeneities associated with As-Se and Ge-Se respond rather independently and a considerable fraction of metastable homopolar bonds present in the atomic fragments are broken and subsequently converted into energetically favored heteropolar bonds. The reduction in As homopolar bonds results in PD [16], i.e.

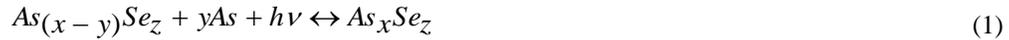

$$As_{(x-y)}Se_z + yAs + h\nu \leftrightarrow As_xSe_z \qquad (1)$$

In the case of Ge atoms, surface oxidation and intrinsic structural changes described by the following chemical reaction lead to photobleaching [10].

$$GeSe_a + Se_b + h\nu \leftrightarrow GeSe_{(a+b)} \quad \text{where } a < 2 \leq a+b \tag{2}$$

Thus there exist two parallel mechanisms of PD and PB in these samples with light illumination. Between them PD is the faster process which saturates rapidly. It is then followed by the slower PB, which requires prolonged illumination to saturate. As a result, we observed an unusual coexistence of PD and PB. To model their reaction kinetics, we used a combination of stretched exponential functions that describe PD and PB separately. Rate equation determining PD can be written as

$$\Delta T = C[\exp\{-(\frac{t}{\tau_d})^{\beta_d}\}] + \Delta T_{Sd} \tag{3}$$

and that for PB

$$\Delta T = \Delta T_{Sb}[1 - \exp\{-(\frac{t}{\tau_b})^{\beta_b}\}] \tag{4}$$

where the subscript 'd' and 'b' corresponds to PD and PB. $\Delta T_S$, $\tau$, $\beta$, t and C are metastable part, effective time constant, dispersion parameter and illumination time and temperature dependent quantity which is equal to maximum transient changes respectively. The net rate equation for the whole process is a summation of respective PD and PB:

$$\Delta T = C[\exp\{-(\frac{t}{\tau_d})^{\beta_d}\}] + \Delta T_{Sd} + \Delta T_{Sb}[1 - \exp\{-(\frac{t}{\tau_b})^{\beta_b}\}] \tag{5}$$

The experimental data fit very well to the stretched exponential functional forms as described in eq. (5) - see fig. 1(b) and (d). Fitting parameters calculated from theoretical fit are listed in Table 1. Note that the effective reaction time for PD is relatively small and is of the order of a few seconds for 2W/cm$^2$. By contrast, PB is a slow process with much longer reaction times compared to that for PD. Interestingly, as we decrease the intensity by a factor of ten, the magnitude of PD and PB remain about the same, but the kinetics change remarkably. In our experiments, we observed a ten time decrease in kinetics, when the pump beam intensity was reduced from 2 to 0.2 W/cm$^2$. Thus the intensity of the pump beam has a predominant role in determining the kinetics of PD and PB.

As the compositional heterogeneities associated with As-Se and Ge-Se fragments appear to respond to light rather independently, we focus on the reversibility of transient effects in $Ge_{19}As_{21}Se_{60}$ film by turning off the pump beam after the complete saturation of PB. On turning off the pump beam, transmission increased further and saturated quickly (fig. 2). When the illumination was switched on subsequently, transmission decreased and reached the values before the illumination was switched off. Subsequent on-off cycles of the pump beam showed reversible transient photodarkening (TPD) every time. The TPD component in $Ge_{19}As_{21}Se_{60}$ is probably of the same origin (in bond switching and atom movement) that was observed in As chalcogenide films and more recently in $Ge_2Se_3$ and $GeSe_2$ [9, 10].

Interestingly, we have measured reversible TPD only after complete saturation of PB and as a result it is of quite interest to check its existence when the film undergoes PD in the initial few seconds. Such study will give information of TPD whether it is a process with time lag similar to PB or instantaneous like that of PD. Recall that in our experiments we observed an initial fast PD which is followed by a slow PB. In that context, we illuminated the sample for a few seconds and turned off the pump beam. Interestingly, after turning off the pump beam, transparency of the film increases, which grows gradually and saturates quickly (fig. 3), however never come back to the initial value prior to illumination (contributions from PD and reversible TPD). As our next step, we turned on the pump beam again and saw the transmission of the film quickly revert back to the initial value.

**Table 1:** Fitting parameters obtained from Eq. 5 that corresponds to PD and PB at two intensities. The subscript b and d refer to bleaching and darkening, respectively.

| Intensity (W/cm$^2$) | Wavelength (nm) | $\tau_d$ (sec) | $\beta_d$ | $\Delta T_{Sd}$ | $\tau_b$ (sec) | $\beta_b$ | $\Delta T_{Sb}$ |
|---|---|---|---|---|---|---|---|
| 2 | 600 | 18 | 0.83 | 0.890 | 650 | 0.64 | 0.187 |
|  | 590 | 16 | 0.79 | 0.910 | 600 | 0.74 | 0.166 |
|  | 580 | 16 | 0.75 | 0.905 | 563 | 0.72 | 0.158 |
|  | 570 | 15 | 0.75 | 0.916 | 547 | 0.74 | 0.169 |
| 0.2 | 600 | 278 | 0.63 | 0.94 | 5780 | 0.73 | 0.158 |
|  | 590 | 242 | 0.66 | 0.95 | 5973 | 0.85 | 0.163 |
|  | 580 | 219 | 0.73 | 0.95 | 5733 | 0.85 | 0.163 |
|  | 570 | 224 | 0.58 | 0.935 | 5802 | 0.74 | 0.213 |

Thus our experiments clearly demonstrate that reversible TPD is an instantaneous process similar to PD. However, when such cycles were repeated with prolonged pump beam illumination, the dynamics changed appreciably. For each long illumination periods there exists a competition between PD (both metastable and transient) and PB. At short times, PB is very weak and therefore not observable in the presence of large PD. At long times PD has saturated (although it includes reversible transient component) and PB begins to increase eventually dominating the response. All our data clearly indicate PD is an almost instantaneous process with light illumination and PB is a slow process.

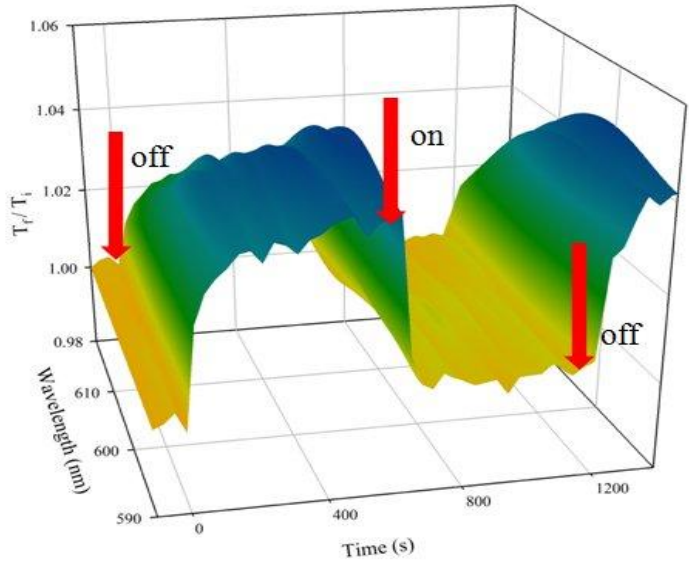

**Fig. 2.** Shows the existence of reversible TPD. When the pump beam was turned off, transparency of the film increases and saturates quickly. However, when we turned on the pump beam, transmission of the film decreases and quickly returned to the pre-switching value. Thus our experiments clearly demonstrate TPD.

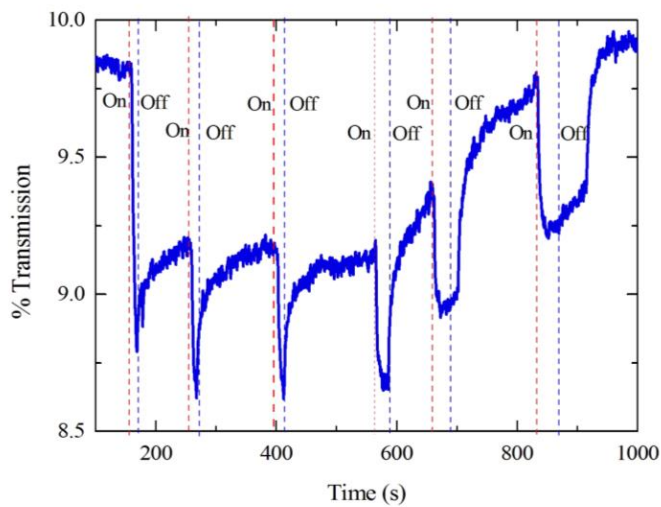

**Fig. 3.** TPD measurements in $Ge_{19}As_{21}Se_{60}$ thin film when it undergoes initial PD with pump beam illumination. The dotted red and blue lines indicate the time at which the pump beam was turned on and off respectively. With pump beam illumination transmission decreases quickly. However, when the pump beam is turned off, transmission increases, but never recovers completely showing the metastable and transient component of PD. For consecutive on and off cycle for short time, the TPD is fully reversible. However, when the sample was illuminated for a long time, sample begins to show PB.

In conclusion, we have demonstrated experimentally the unusual coexistence of fast PD and slow PB in $Ge_{19}As_{21}Se_{60}$ thin films, when illuminated with DPSSL of $\lambda = 671$nm. The observed effect is explained by assuming that with illumination As-Se and Ge-Se compositional heterogeneities respond rather independently and give rise to PD and PB respectively. Notably, the kinetic curves of both PD and PB follow stretched exponential response. Moreover PD is a fast process that appears to begin almost instantaneously with light illumination and saturates in tens of seconds. On the other hand, PB is a slower process and saturates only after prolonged illumination. Apart from PD/PB, the sample also shows transient/reversible effects similar to those observed in As-based CG and $GeSe_2$ thin films.

**Acknowledgements**

The authors thank Department of Science and Technology (Project no: SR/S2/LOP-003/2010) for financial support. They also thank the National Science Foundation for supporting our international collaboration through International Materials Institute for New Functionality in Glass (DMR-0844014).